\documentclass[a4paper]{article}
\usepackage[utf8x]{inputenc}
\usepackage[english]{babel}
\usepackage{cite}
\usepackage{wrapfig}
\usepackage{graphicx}
\usepackage{amssymb}
\usepackage{amsfonts}
\usepackage{amsmath}
\usepackage{longtable}
\usepackage{rotating}
\usepackage{lscape}
\usepackage{epsfig}
\usepackage{multirow}

\usepackage{float}

\begin{document}

	\title{Study of strongly intense quantities and robust variances in multi-particle production at LHC energies}
	\maketitle
	\author{S.\,Belokurova\footnote{E-mail: sveta.1596@mail.ru}}

	\begin{abstract}		
		The strongly intense quantities and robust variances in processes of multi-particle production in pp and AA interactions at LHC energies was studied. The Monte Carlo and analytic modelling of these quantities in the framework of a quark-gluon string model were implies. The string fusion effects were also taken into account by implementing of a lattice (grid) in the impact parameter plane. Strongly intensive variable $\Sigma(n_F,\ n_B)$ was calculated for different energies for two values of the width of the observation rapidity windows as a function of the distance between the centres of this windows.
		Scaled variance $\omega_n$ and robust variance $R_n$ for different energies and for different width of the observation rapidity window was calculated by MC simulations. 
	\end{abstract}
	\vspace*{6pt}
	
	
	\label{sec:intro}
	\section*{Introduction}	
	As is well known, at the present the quantum chromodynamics does not enable to describe numerically the soft part of multi-particle production. The different versions of the QCD-inspired quark-gluon string model are used for a description of this component of hadronic interaction at high energy. One of the most popular approaches is the string model \cite{col1, col2, col3, col4}. In this model at first stage the color quark-gluon strings are formed. At second stage the hadronization of these strings produces the observed hadrons. 

	\section*{Generation of the string configuraton}
	At first we formulate the MC algorithm based on the string model. In our approach we will consider that each cut pomeron corresponds to formation of two strings \cite{col4}. To take into account string fusion one should know not only the mean number of pomerons in $pp$ collisions at a given impact parameter $b$, but also the event by event distribution of the number of pomerons around this mean value.
	This distribution at a given value of the impact parameter $b$ at $N\geq 1$was chosen in the following form:
	\begin{equation}
	\label{pomeron distribution} \widetilde{P}(N,b) = {P(N,b)}/ [1 - P(0,b)],
	\end{equation}
	where $P(N,b)$ is the poissonian distribution with some parameter $\overline{N}(b)$:
	\begin{equation}
	\label{poisson} P(N,b) = e^{-\overline{N}(b)} {\overline{N}(b)^{N}}/ {N!}\ \ .
	\end{equation}
	 The difference of our distribution $\widetilde{P}(N,b)$ (\ref{pomeron distribution}) from the poissonian one (\ref{poisson}) is only in excluding of the case $N=0$: $\widetilde{P}(0,b)=0$,	which corresponds to the absence of the non-diffractive scattering.	
	
	According to \cite{PoS2012}, we suppose that in the proton-proton collision at the impact parameter $b$ the string density in transverse plane at a point $\vec{s}$ is proportional to
	\begin{equation}
	\label{strind density}
	w_{str}(\vec{s}, \vec{b}) \sim \frac{1}{\sigma_{\!pp}(b)}
	T(\vec{s}-\vec{b}/2)T(\vec{s}+\vec{b}/2) \ ,
	\end{equation}
	where the $T(\vec{s})$  is the partonic profile function of nucleon.
	We will use	for the partonic profile function of nucleon the simplest gaussian distribution:
	\begin{equation}
	\label{a6}
	T(s) =
	\frac
	{e^{-s^2/\alpha^2}}
	{\pi\alpha^2}\ .
	\end{equation}
	Substituting (\ref{a6}) in (\ref{strind density}) one gets
	\begin{equation}
	\label{factor}
	w_{str}(\vec{s}, \vec{b}) \sim
	\frac{1}{\sigma_{\!pp}(b)}
	e^{-2 s^2 / \alpha^2}
	e^{-b^2 / 2\alpha^2}  \ .
	\end{equation}
	Simultaneously in this approach we have $\overline{N} (b)=N_0 e^{-b^2/2\alpha^2}$, where the parameter $N_0$ depends on initial energy. 
	
	As has been shown in \cite{PoS2012}, in the framework of this assumptions the average number of pomerons $	\left\langle N_{pom}(E)\right\rangle$, the scaled variance of number of pomerons $	\omega_{N_{pom}}(E)$, the cross-section of non-diffractive $pp$ interaction $	\sigma_{\!pp} $ and the probability $P(N)$ to have $N$ cut pomerons in a non-diffractive \textit{pp} collision has the following form:
	\begin{equation}
	\label{N pom}
	\left\langle N_{pom}(E)\right\rangle  = \frac{ N_0}{ E_1( N_{0}) + \gamma + \ln{N_0}} \ ,\ 
	E_1(x) = \int_{x}^{\infty}\frac{e^{-t}}{t}dt,
	\end{equation}
	\begin{equation}
	\label{omNpom}
	\omega_{N_{pom}}(E)= 1+\frac{N_0}{2} - \left\langle N\right\rangle _{pom}(E) \ ,
	\end{equation}
	\begin{equation}
	\label{ND cross section}
	\sigma_{\!pp} = 2\pi \alpha^2 \left[  E_1( N_{0}) + \gamma + \ln{N_0}\right] ,
	\end{equation}	
	\begin{equation}
	\label{P pom}
	P(N) =
	\frac{2\pi\alpha^2}{\sigma_{\!pp}^{} N}
	\left[1- e^{-N_0} \sum_{l=0}^{N-1} N_0^l/l! \right] .
	\end{equation}
	
	The last formula shows that this approach is equivalent to the Gribov-Regge approach, as it was noted in \cite{PoS2012}.
	This enables to connect  the parameters $N_0$ and $\alpha$ of string fusion model, which describe the dependence of the mean number of pomerons on the impact parameter $b$ with  the parameters of the pomeron trajectory and its couplings to hadrons:
	\begin{equation}
	\label{paramfix}
	\alpha=\sqrt{\frac{2\lambda}{C}}/5.05\  fm,\  
	N_0=\frac{2\gamma_{pp} C}{\lambda} \exp(\Delta\xi)  ,\  
	\lambda=R_{pp}^2+\alpha'\xi,\ \xi=\ln (s/1 GeV^2)\ .
	\end{equation}
	Here  $\Delta$ and $\alpha'$ are the intercept and the slope of the
	pomeron trajectory.
	The parameters $\gamma$ and $R_{pp}$ characterize the coupling of the pomeron trajectory
	with the initial hadrons. The quasi-eikonal parameter $C$ is related to
	the small-mass diffraction dissociation of incoming hadrons.
	
	For the case of $pp$ collisions the following numerical values of the parameters were chosen to describe the multiplicity and the non-diffractive cross section:
	\begin{equation}
	\label{ourparam}
	\Delta= 0.2,\ \ \ \alpha'=0.05\ GeV^{-2},\ \ \ \gamma_{pp}=1.035\ GeV^{-2},\
	\ \  R_{pp}^2=3.3\ GeV^{-2},\ \ \ C=1.5.
	\end{equation}	
	
	The string density in the transverse plane increases with the growth of initial energy or going to collisions of nuclei, and it is necessary also to take into account an interaction
	between the strings \cite{int1, int2}, which leads to the formation of fused strings \cite{model1, model2}.
	To simplify the account of string fusion processes in our calculations we use the version of the model with the transverse lattice (grid) \cite{uni distr 0,uni distr 1,uni distr 2}. In the model the transverse plane is divided into cells, which area is equal to the string transverse cross-section. It is supposed that the strings with the centers in the same cell are fused.

	In the framework of the string fusion model \cite{model1, model2} the dependence of the average number of particles formed from decay of the fused strings in the cell on the number of strings, $\eta_i$, in the rapidity observation window of width	$ \delta y $ have the following form:
	\begin{equation}
		\label{mean mult}
	\overline{n}\left( \eta_i\right)=\mu_0 \delta y \sqrt{\eta_i },
	\end{equation}
	where $\mu_0$ is the average number of a particles produced from the hadronizations of the one string in the window of width	$ \delta y = 1 $.
	In our calculations the following numerical value of the string radius was chosen:
	\begin{equation}\label{r str}
	r_{str} = 0.2\ fm.
	\end{equation}	

	We assume that the number of particles produced from the hadronizations of the strings in i-th cell in the rapidity observation window of width $ \delta y $ is distributed over the negative binomial distribution (NBD) with mean value \eqref{mean mult} and scaled variance:
	\begin{equation}
	\omega_\mu (\delta y,\ \eta) = 1+ \delta y  \mu_0^\eta  J_{FF}^{\eta}, 
	\end{equation}	
	see the paper \cite{NPA15}, where 
	\begin{equation} 
	J_{FF}^{\eta} = \frac{1}{(\delta y_F)^2} \int_{\delta y_F} dy_1 \int_{\delta y_F} dy_2 \Lambda_\eta (y_1 - y_2)
\end{equation}
	and $\Lambda_\eta (\Delta y )$ is the two-particle (pair) correlation function, which was chosen in the simplest form
	\begin{equation}
	\label{corr function}
	\Lambda_\eta (\Delta y ) = \Lambda_0^\eta e^{-\frac{|\Delta y|}{y_{corr}^\eta}},
	\end{equation}
	$ y_{corr}^\eta$ is a characteristic correlation length in the rapidity space. 
	In accordance with the physical picture of the string fusion in the model we assume that the dependence of the parameters on the string density, $\eta$, is as follows
	\begin{equation}
	y_{corr}^\eta = \frac{y_1}{\sqrt{\eta}},\ 
	\mu_0^\eta = \mu_0 \sqrt{\eta}.
	\end{equation}	

	For the correlation function chosen in the simplest form, \eqref{corr function}, the integral $J_{FF}^{\eta}$ can be calculated explicitly:
\begin{equation}
	J_{FF}^{\eta}
	=\frac{2\Lambda_0^{\eta}}{(\delta y)^2} y_{corr}^{\eta} \left( \delta y-y_{corr}^{\eta}\left( 1-e^{-\frac{ \delta y}{y_{corr}^{\eta}}}\right)  \right).
\end{equation}
	
	Parameters $y_1$ and $ \Lambda_0^\eta$ was chosen to obtain a correspondence with the results for $\Sigma(n_F,\ n_B)$ obtained in \cite{Vechernin18, Andronov2019, Vechernin19} using the pair correlation function extracted in \cite{NPA15} in the approximation of identical strings from ALICE \cite{ALICE15} experimental data. The value of the parameter $ \mu_0 $ was chosen to describe $dN/dy$ distribution at different energies taken from \cite{Pogosayn12rep, TOTEM, CMS, ALICE7n, TOTEM13}:
		\begin{equation}
	\mu_0 = 0.7,\ \ 
	y_1 = 2.7,\ \ 
	\Lambda_0^\eta = 0.8.
	\end{equation}

Based on the foregoing, the MC algorithm was elaborated.  The developed algorithm was used to generate the events at the following energies: 60 Gev, 900 Gev, 7 TeV, 13 TeV.

	\section*{Calculation of the $\Sigma(n_F,\ n_B)$, $\omega_n$, $R_n$}
	The definitions of of the strongly intensive variable $\Sigma(n_F,\ n_B)$, scaled variance $\omega_n$ and robust variance $R_n$ are as follows \cite{Gorenstein11,AndronovTMPh15}:
	\begin{equation} \label{sigma}
	\Sigma(n_F,\ n_B) \equiv \frac{ \left\langle n_F\right\rangle\omega_{n_B} +\left\langle n_B\right\rangle\omega_{n_F} - 2 cov (n_F,\ n_B) }{  \left\langle n_F\right\rangle+ \left\langle n_B\right\rangle },
	\end{equation}
	\begin{equation}\label{definitions}
	\omega_n \equiv \frac{ \left\langle n^2\right\rangle- \left\langle n\right\rangle^2 }{  \left\langle n\right\rangle },\ 
	R_n \equiv \frac{\omega_n - 1 }{ \left\langle n\right\rangle } = \frac{\left\langle n(n+1)\right\rangle }{\left\langle n\right\rangle ^2}-1.
	\end{equation}
	
	As it was shown in \cite{Belokurova Vechernin 2019}, for the strongly intensive variable $\Sigma(n_F,\ n_B)$ \eqref{sigma} 
	the following expression can be obtained:
	\begin{equation}\label{sigma rewritten}
		\Sigma(n_F,\ n_B) = \sum_{\eta = 1}^{\infty} \frac{\left\langle n\right\rangle _\eta}{\left\langle n\right\rangle} \Sigma_\eta (\mu_F,\ \mu_B),\ 
	\Sigma_\eta (\mu_F,\ \mu_B) = 1 + \mu_0^\eta \delta y \left[ J_{FF}^\eta - J_{FB}^\eta\right] ,
	\end{equation}
	where $ \Sigma_\eta (\mu_F,\ \mu_B) $ is the variable $ \Sigma $ for a cell with $\eta$ strings, $ \left\langle n\right\rangle _\eta $ is the average numbers of particles produced from the decay of all string clusters with $\eta$ strings, $ \left\langle n\right\rangle $ --- the multiplicity,	
	\begin{equation} \label{integrals}
	J_{FB}^{\eta} = \frac{1}{\delta y_F \delta y_B} \int_{\delta y_F} dy_1 \int_{\delta y_B} dy_2\ \Lambda_\eta (y_1 - y_2).
	\end{equation}
	For the correlation function of the simplest form \eqref{corr function}, we have
	\begin{equation}\label{J FB}
	J_{FB}^{\eta} 
	=\frac{\Lambda_0^{\eta} \left( y_{corr}^{\eta}\right) ^2}{(\delta y)^2}  e^{\frac{ -\Delta y  }{y_{corr}^{\eta}}} \left( 
	e^{\frac{ \delta y  }{y_{corr}^{\eta}}}
	-2
	+e^{\frac{- \delta y  }{y_{corr}^{\eta}}}
	\right),
	\end{equation}
	where $ \delta y = \delta y_F = \delta y_B $ is the rapidity observation window, $ \Delta y $ is the rapidity distance between the centers of observation windows	(formula \eqref{J FB} was obtained in the case $ \Delta y > \delta y$).
	
	\section*{Results}
	
\begin{figure}[h]
		\centering\
		\begin{tabular}{cc}
			\hspace{-1cm}	\includegraphics[width=0.6\linewidth]{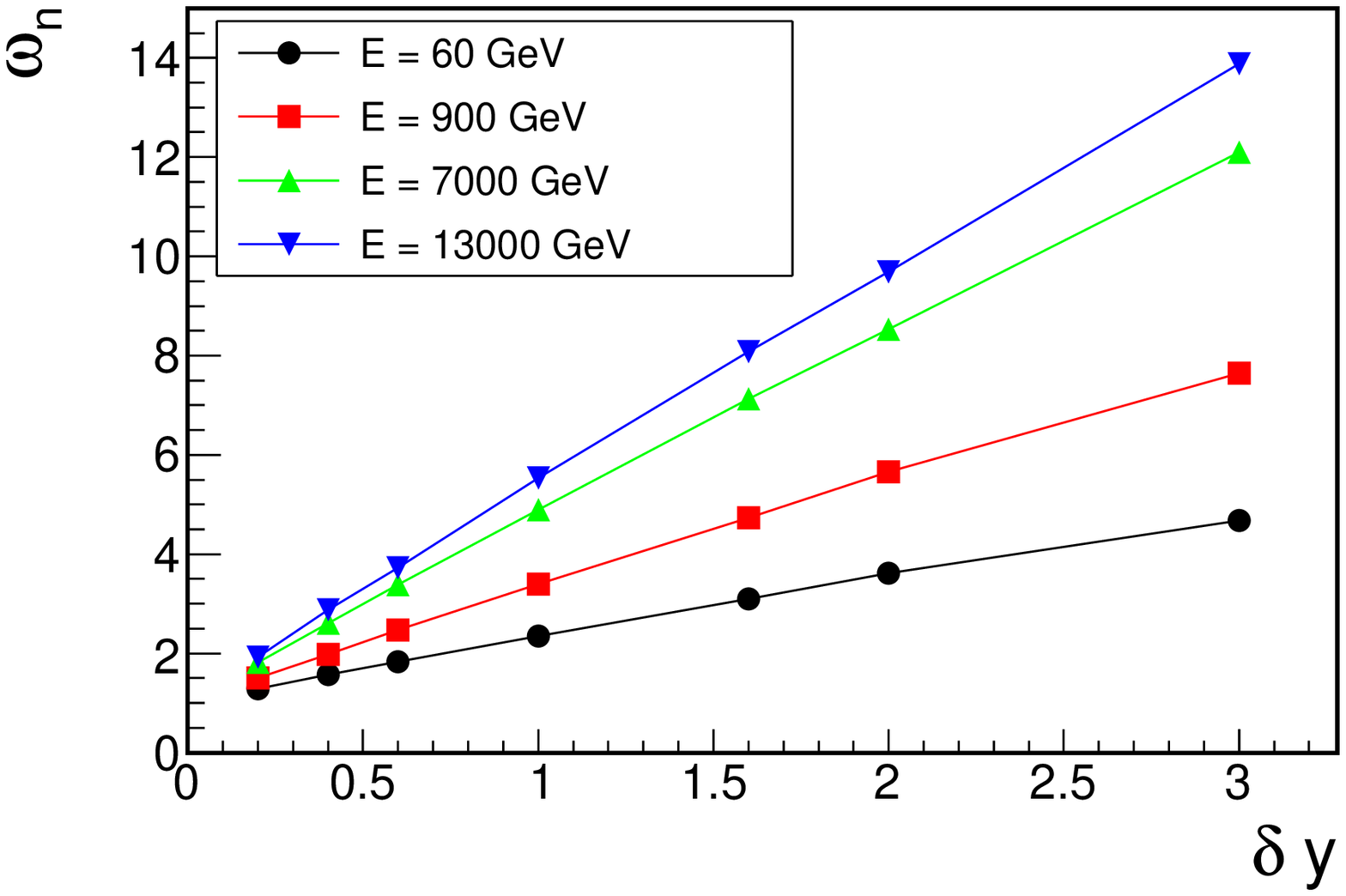}
			&	\hspace{-1.3cm}	\includegraphics[width=0.6\linewidth]{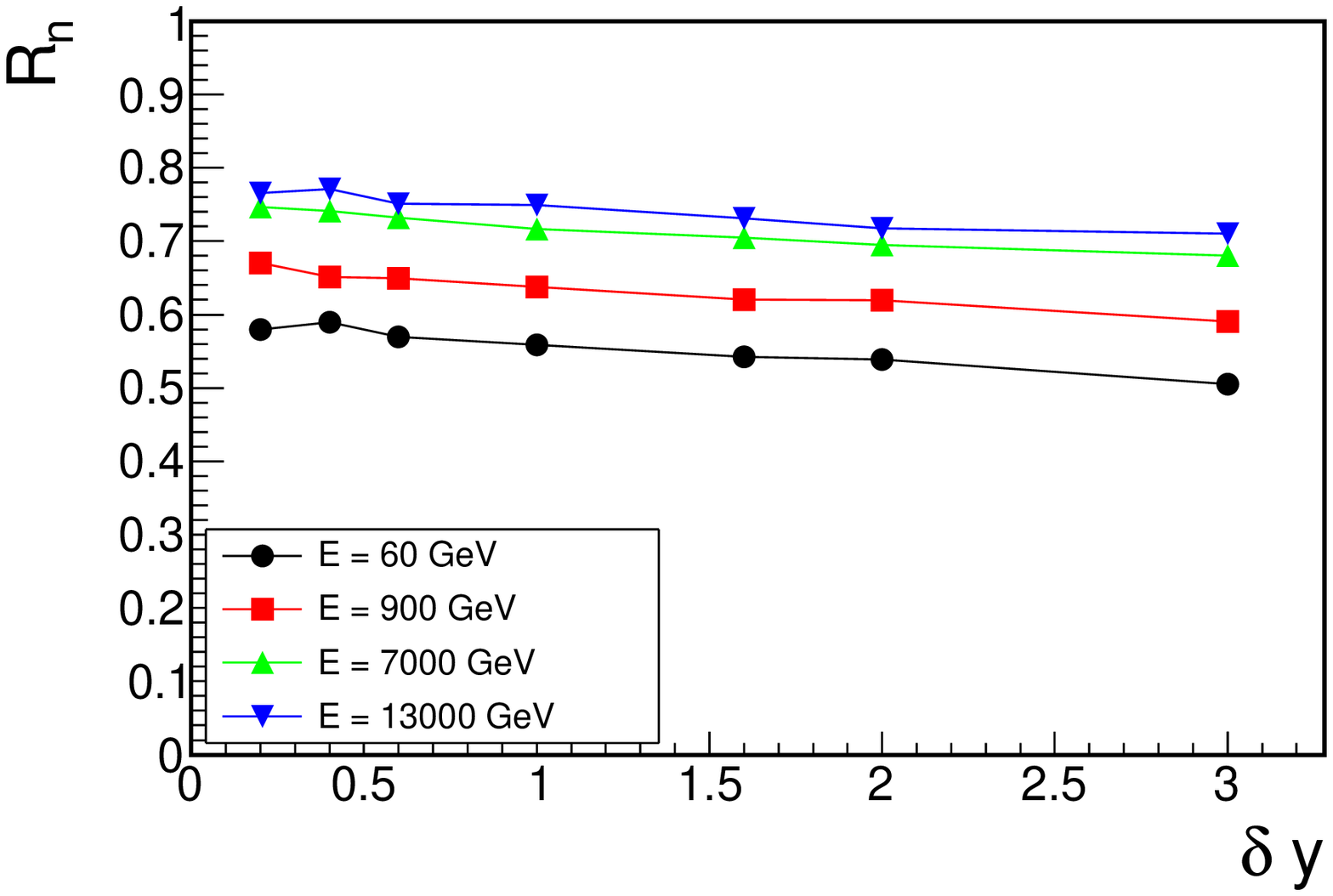}\\
		\end{tabular} 
		\caption{Results for scaled variance $\omega_n$ and robust variance $R_n$ calculated with help of \eqref{definitions} as a function of the rapidity width of the observation window $\delta y$ for min.bias $pp$ interactions at energies 60 - 13000 GeV}
		\label{omegaRn}
	\end{figure}

	For the calculation of the $\Sigma(n_F,\ n_B)$, $\omega_n$ and $R_n$ formulas \eqref{sigma rewritten} and \eqref{definitions} was used. 
	$\omega_n$ and $R_n$ were studied as a function of rapidity width of the observation windows $\delta y$ for min.bias $pp$ interactions at different energies. This dependence is shown in the fig. \ref{omegaRn}.
	
\begin{figure}[h]
		\centering
		\begin{tabular}{cc}
			\hspace{-1cm}	\includegraphics[width=0.6\linewidth]{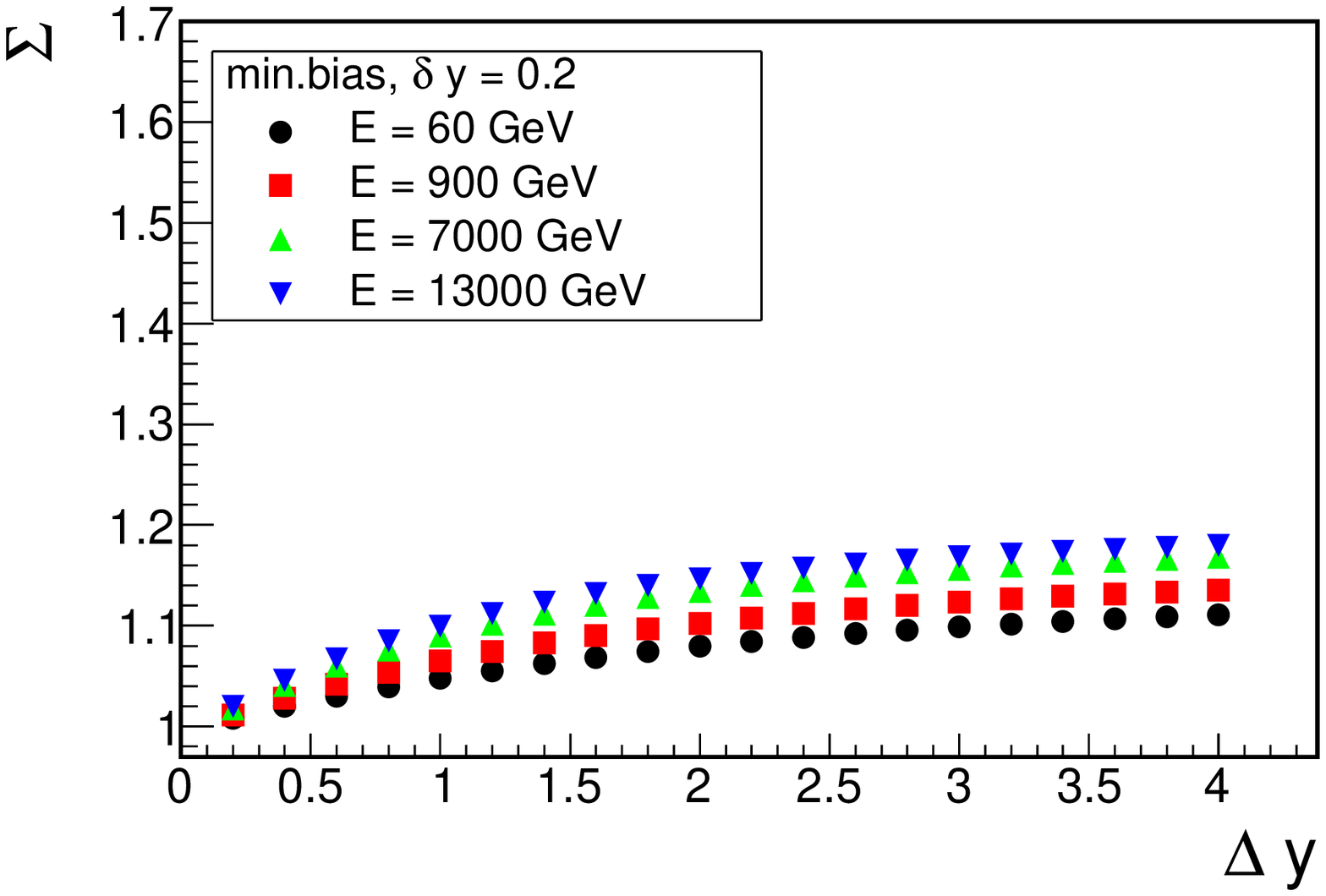}
			&	\hspace{-1.3cm}	\includegraphics[width=0.6\linewidth]{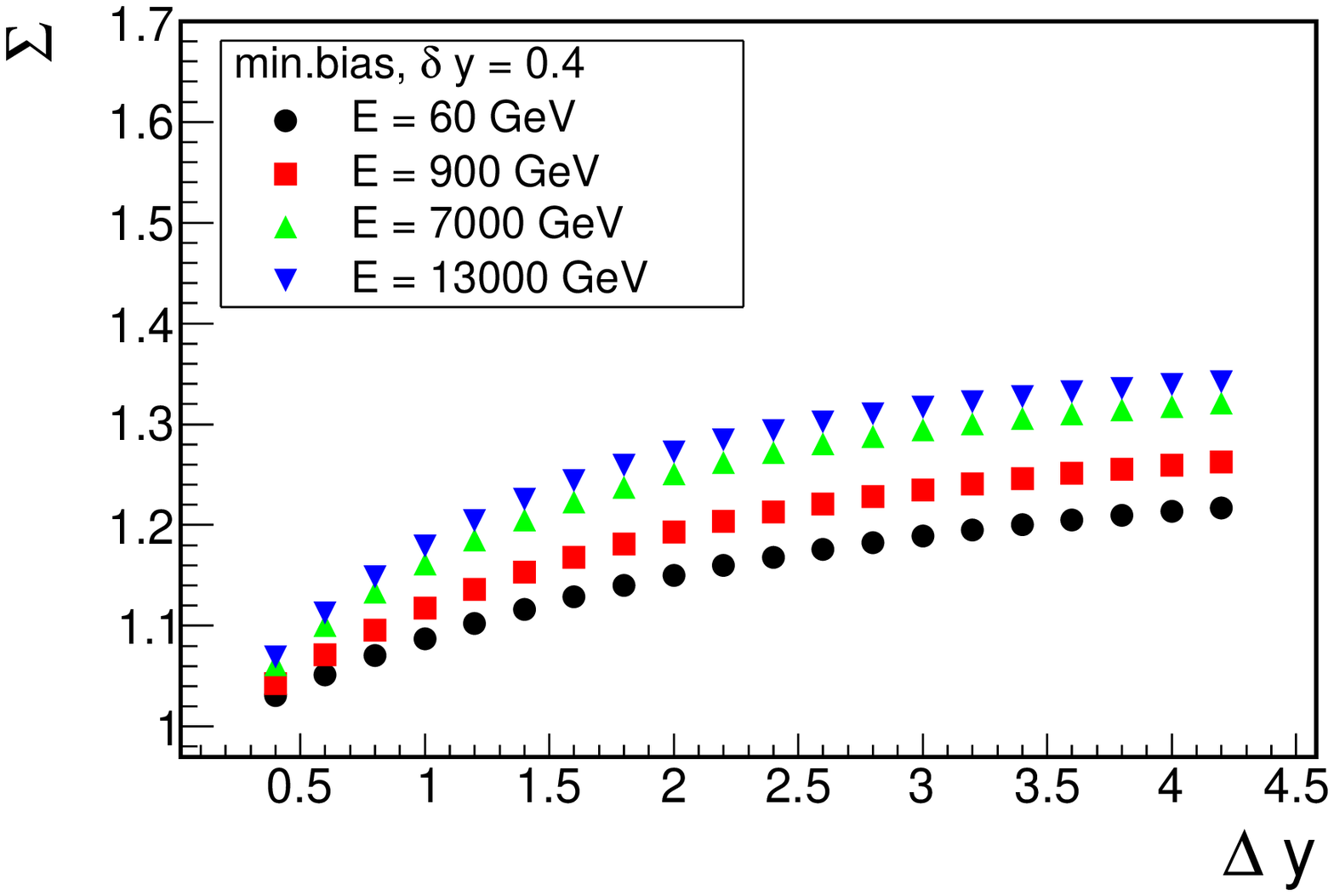}\\
		\end{tabular} 
		\caption{Results for the strongly intensive variable $\Sigma(n_F,\ n_B)$ calculated with help of \eqref{sigma rewritten} as a function of the rapidity distance between the observation windows $\Delta y$ for min.bias $pp$ interactions at energies 60 - 13000 GeV for rapidity width of the observation windows $ \delta y = 0.2$ and $\delta y = 0.4 $.}
		\label{Sigma_minbias}
	\end{figure}

$\Sigma(n_F,\ n_B)$ were studied as a function of the rapidity distance between the observation windows $\Delta y$ for min.bias $pp$ interactions at different energies for rapidity width of the observation windows $ \delta y = 0.2$ and $\delta y = 0.4 $. Results for this study is shown on the fig. \ref{Sigma_minbias}. As one can see $\Sigma(n_F,\ n_B)$ increase with initial energy of pp collision.

	\begin{figure}[H]
	\centering
	\begin{tabular}{cc}
				\hspace{-1cm}	\includegraphics[width=0.5\linewidth]{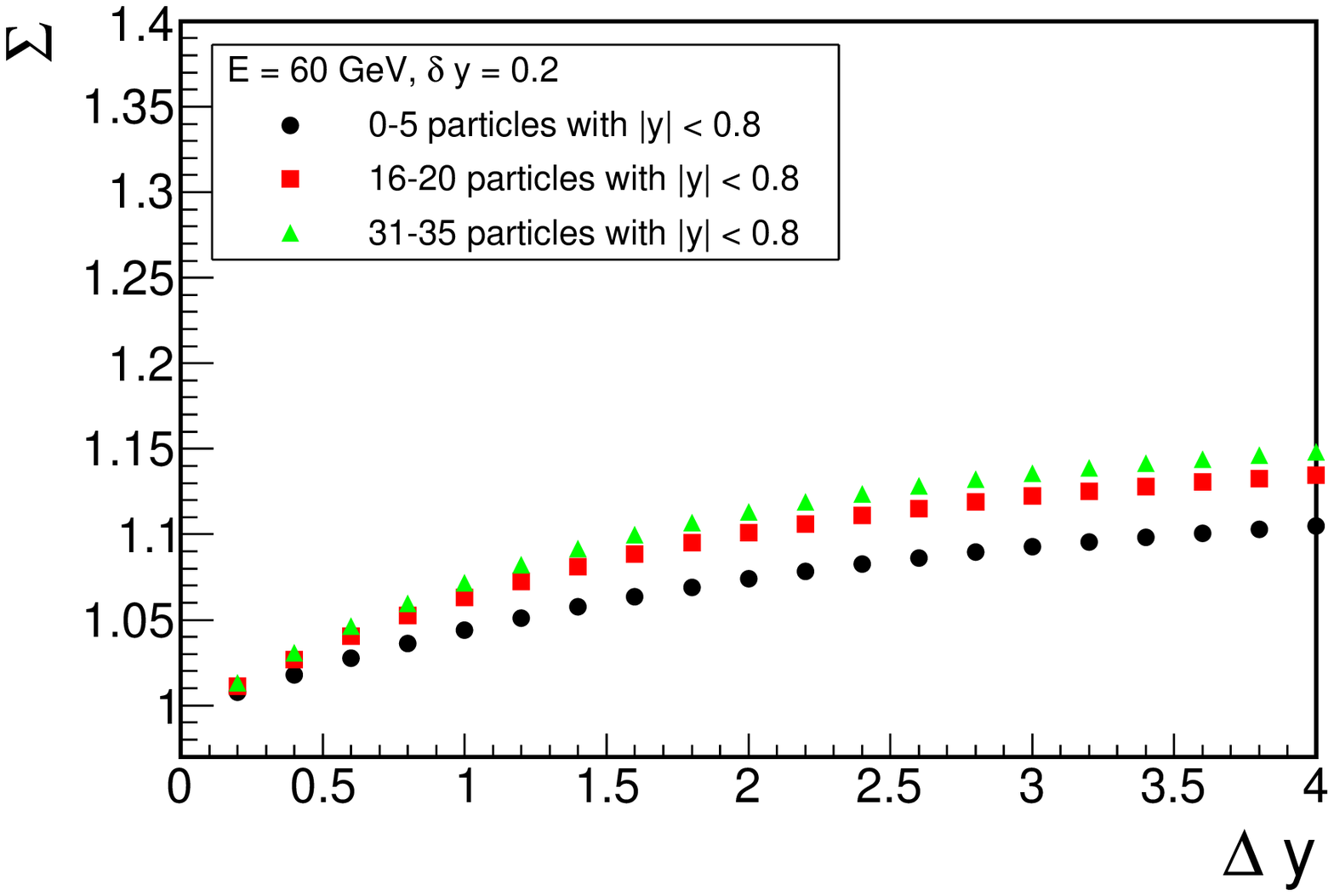}
			&	\hspace{-1.2cm}	\includegraphics[width=0.5\linewidth]{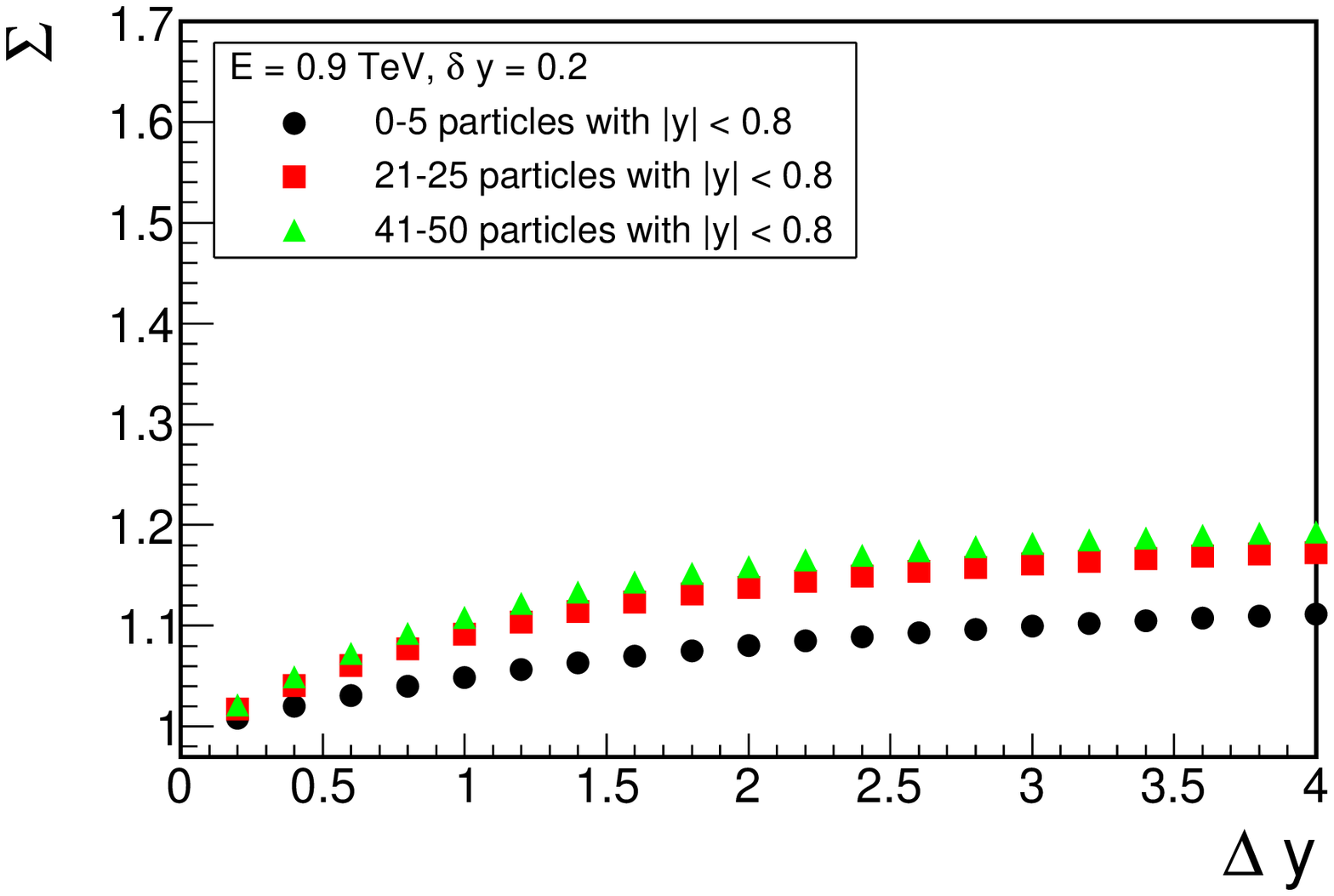}\\
				\hspace{-1cm}	\includegraphics[width=0.5\linewidth]{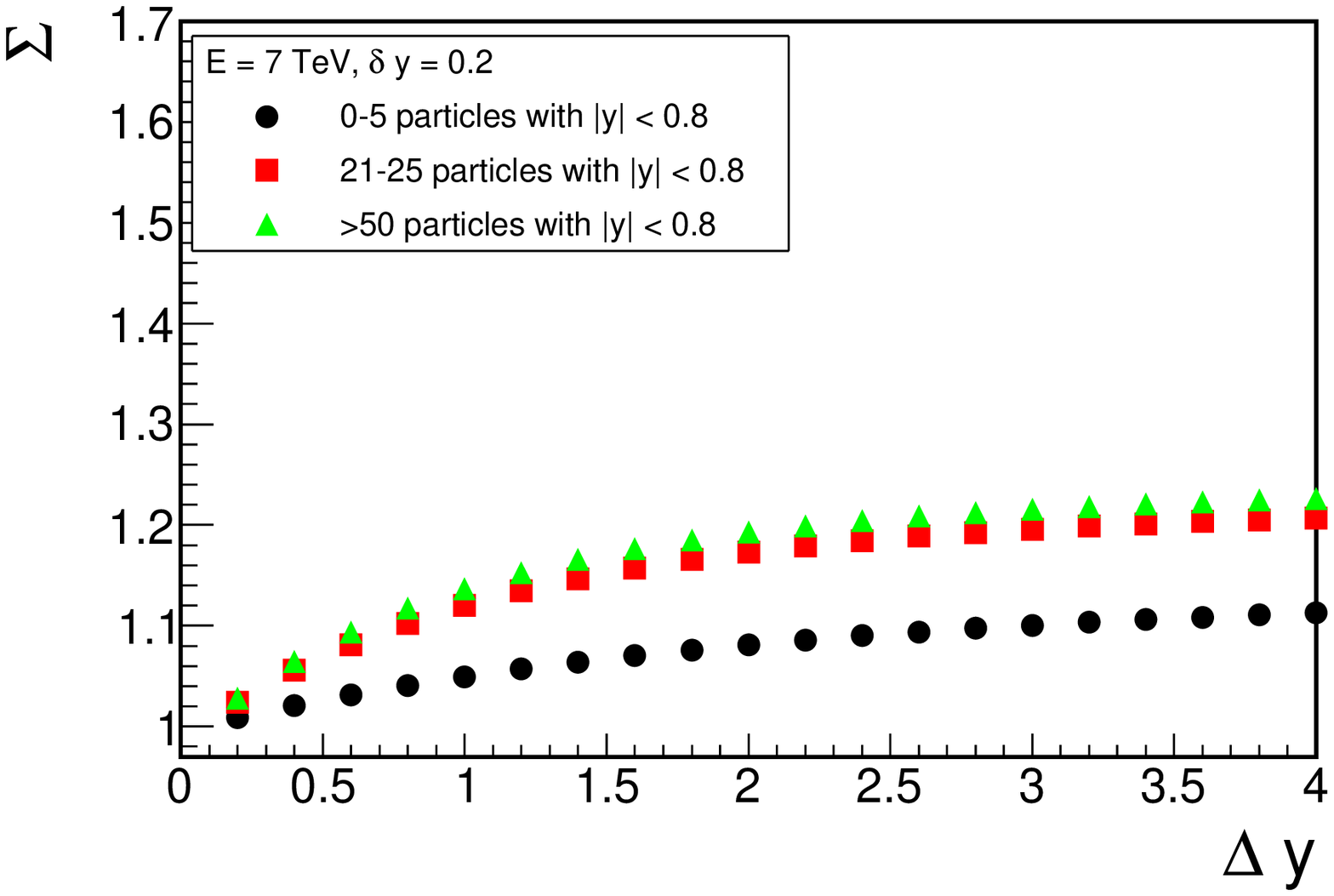}  
			&	\hspace{-1.2cm}	\includegraphics[width=0.5\linewidth]{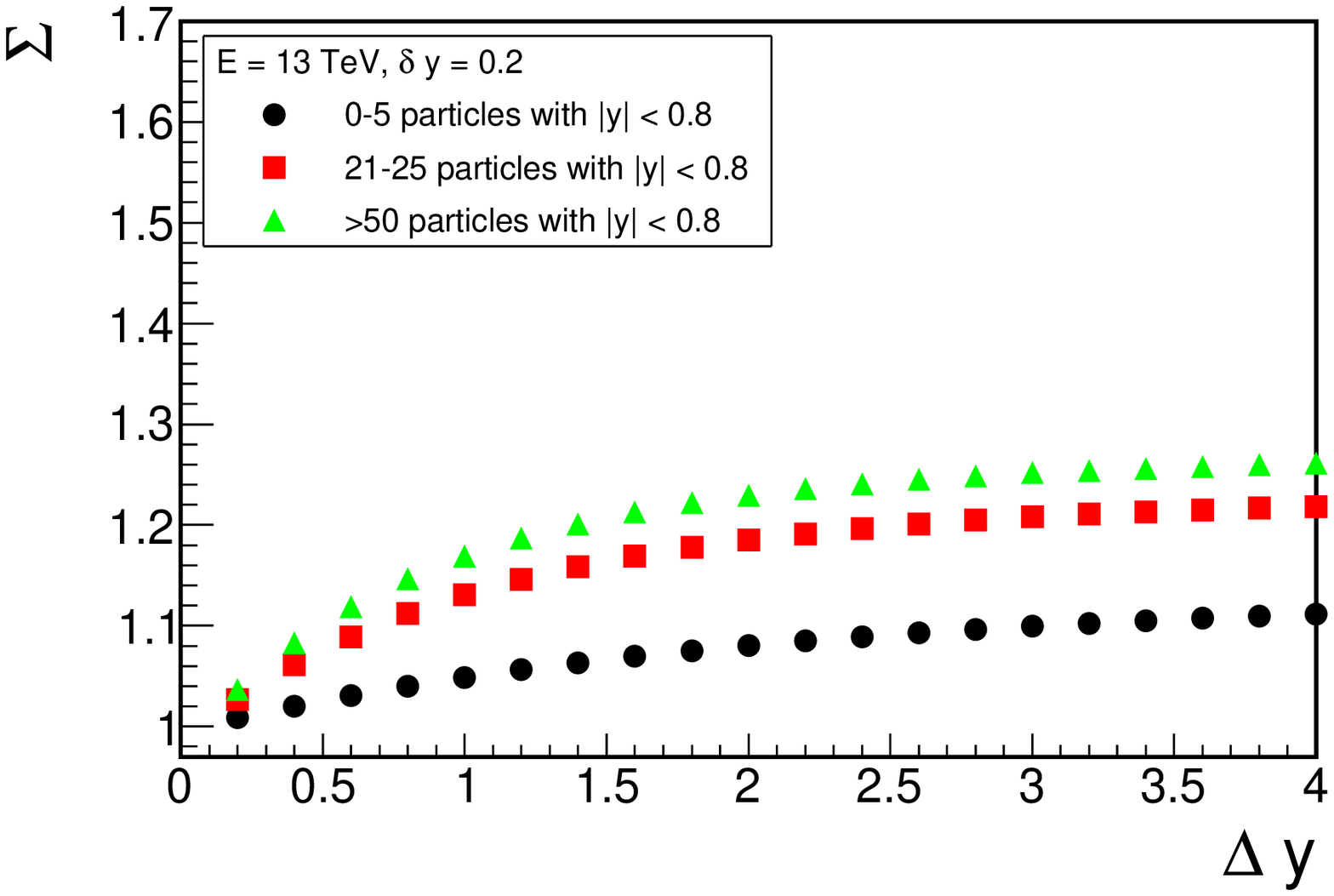}  \\ 
	\end{tabular} 
	\caption{Results for the strongly intensive variable $\Sigma(n_F,\ n_B)$ calculated with help of \eqref{sigma rewritten} as a function of the rapidity distance between the observation windows $\Delta y$ for different centralities of $pp$ interactions at energies 60 - 13000 GeV for rapidity width of the observation windows $ \delta y = 0.2$.}
	\label{Sigma02}
	\end{figure}

	\begin{figure}[H]
	\centering
	\begin{tabular}{cc}
		\hspace{-1cm}	\includegraphics[width=0.5\linewidth]{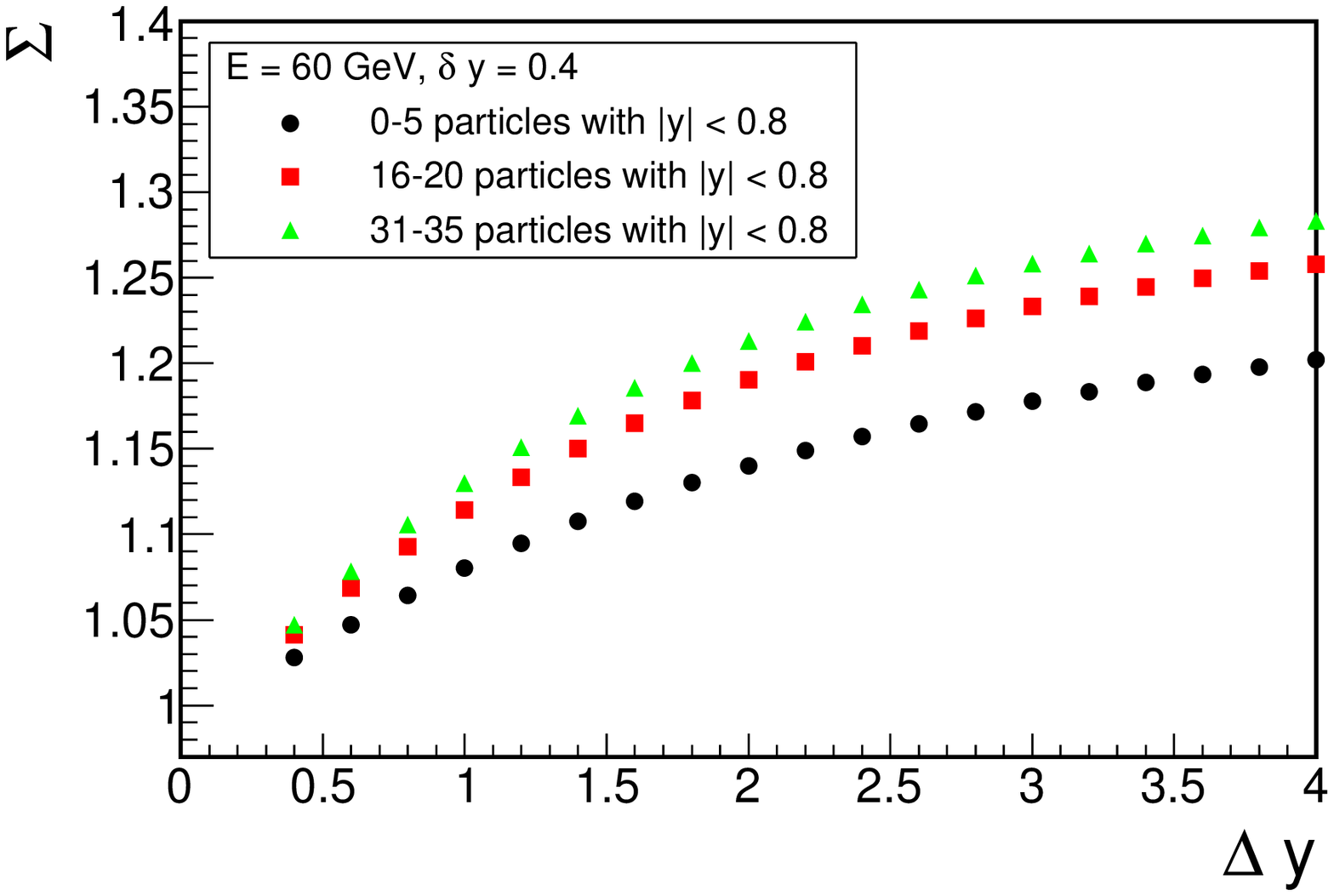} & 
		\hspace{-1.2cm}	\includegraphics[width=0.5\linewidth]{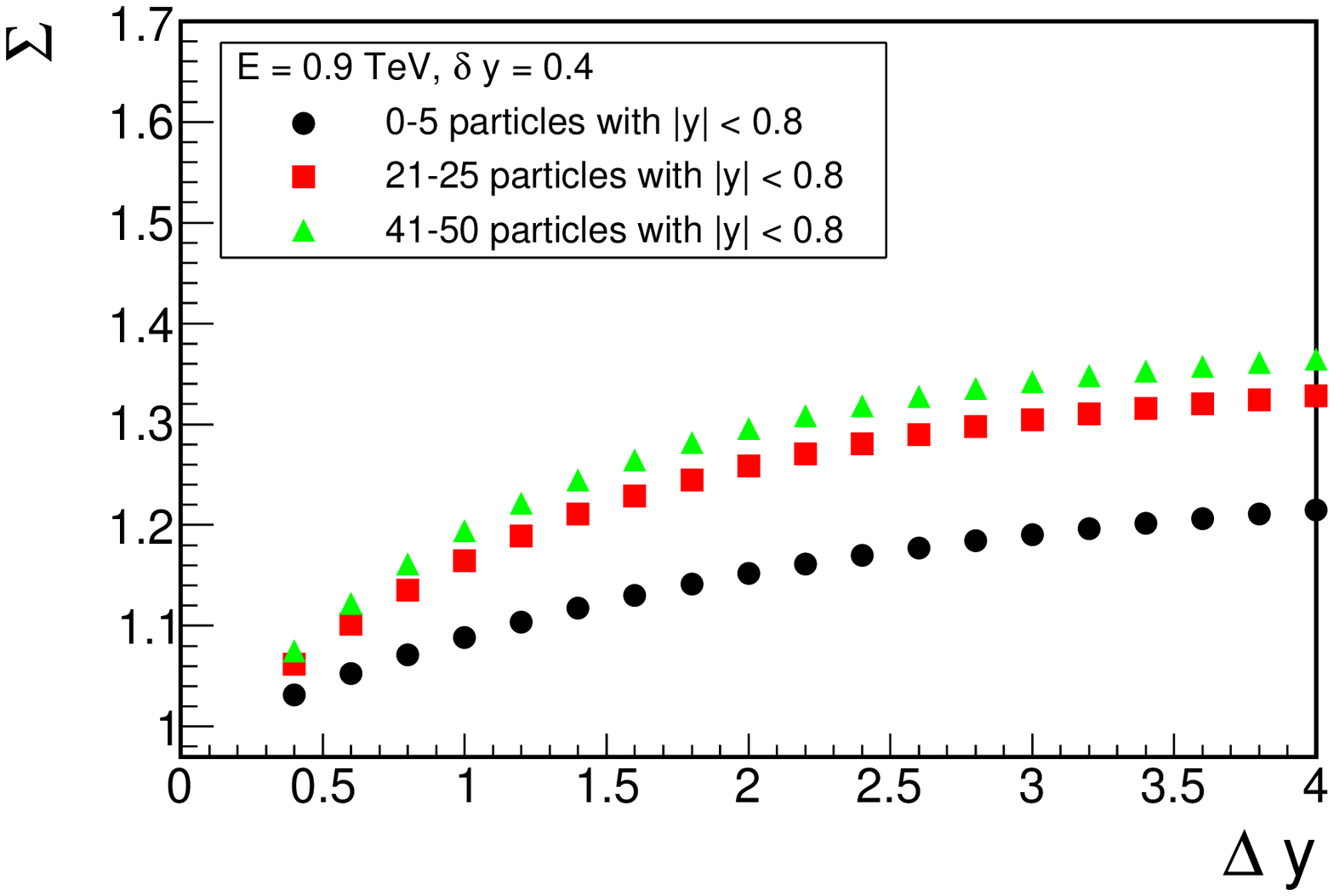} \\
		\hspace{-1cm}	\includegraphics[width=0.5\linewidth]{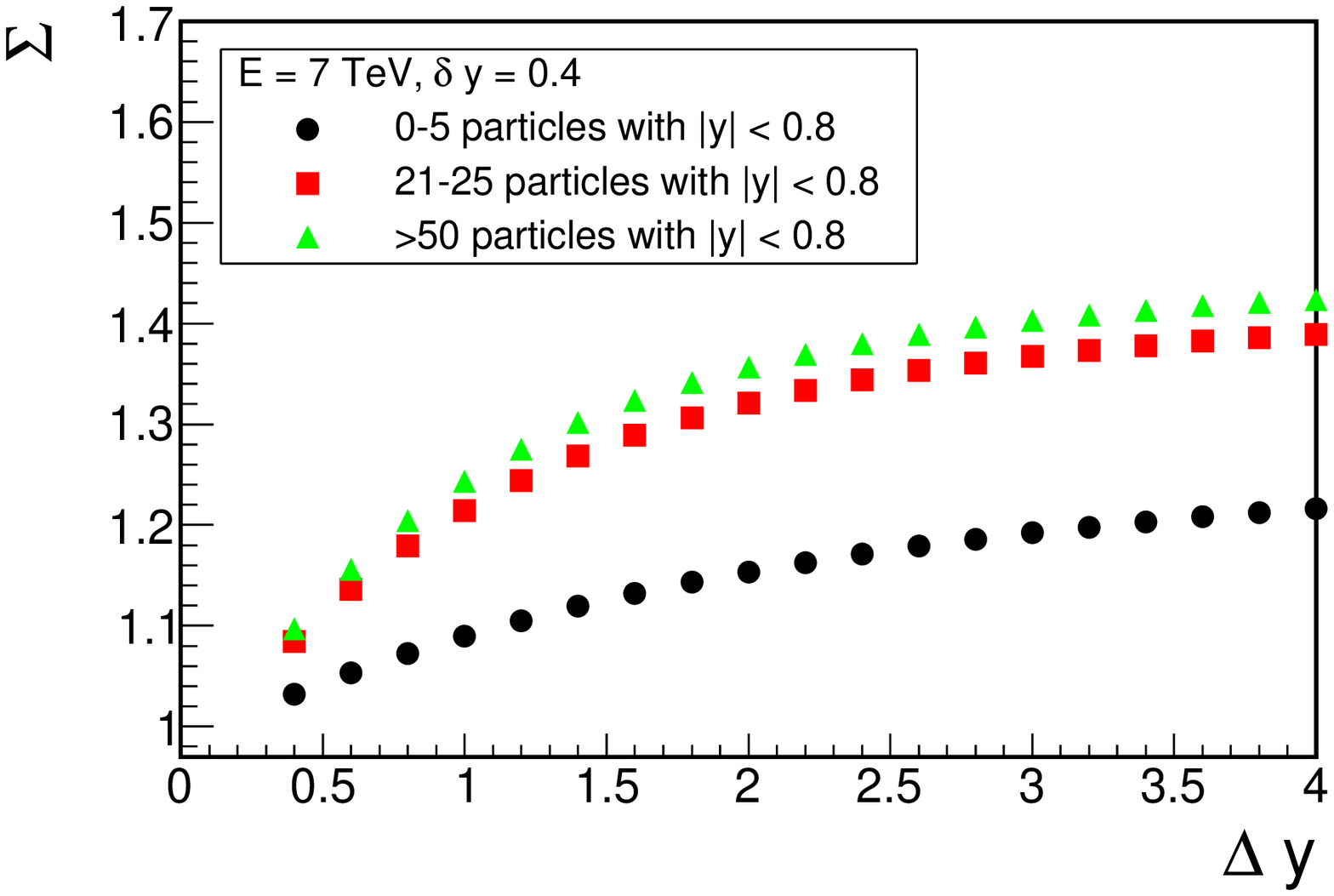}  & 
		\hspace{-1.2cm}	\includegraphics[width=0.5\linewidth]{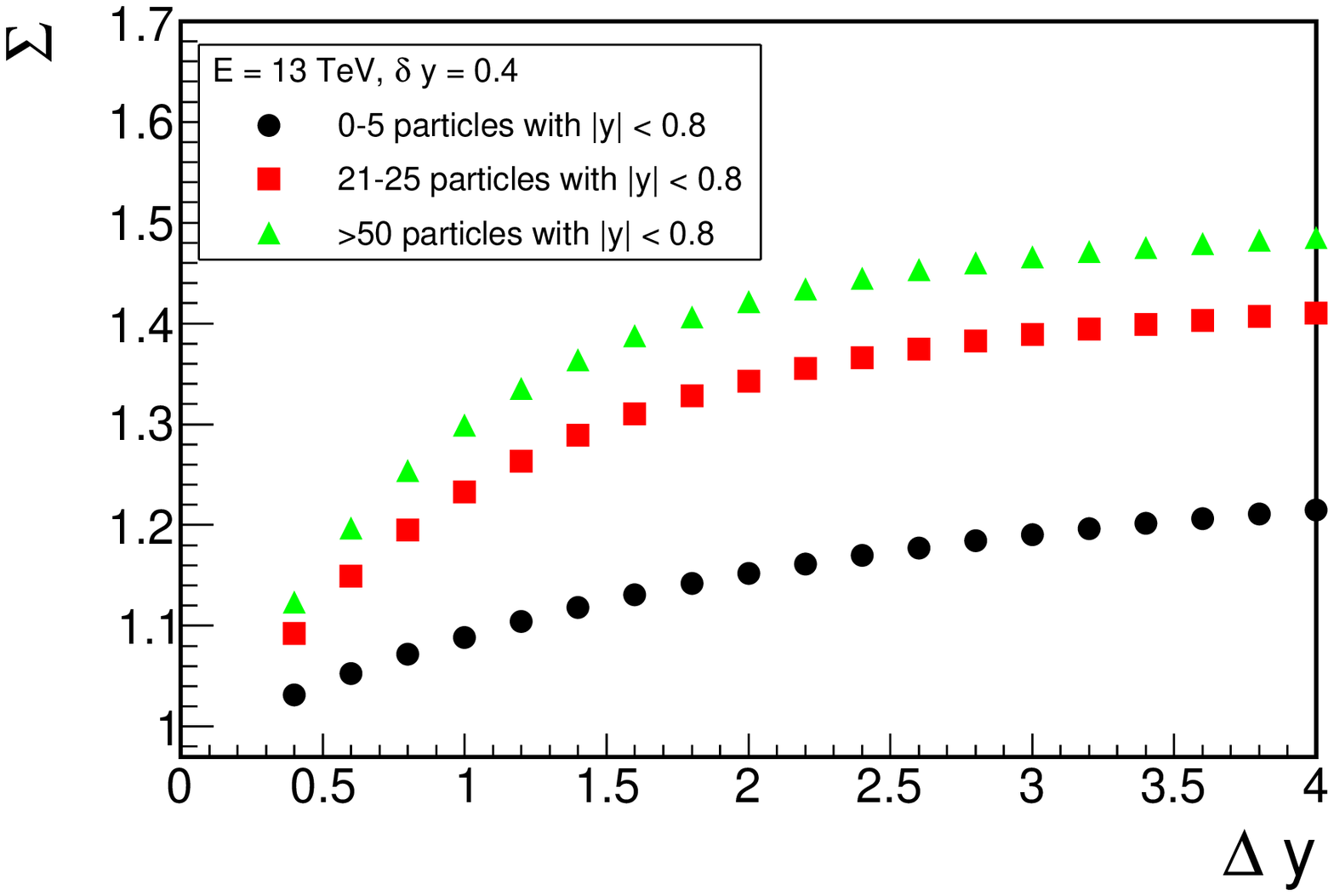}
	\end{tabular} 
	\caption{The same as in fig.\ref{Sigma02} but for rapidity width of the observation windows $ \delta y = 0.4$.}
	\label{Sigma04}
	\end{figure}

The dependence of the strongly intensive variable  $\Sigma(n_F,\ n_B)$ on the collision centrality was also investigated at different energies. As one can see in the figs.  \ref{Sigma02}, \ref{Sigma04}  $\Sigma(n_F,\ n_B)$ grows with increasing centrality of the pp collision.

	\section*{Summary}
	
		The quark-gluon string model approach and the MC algorithm for the analysis of high energy pp collisions were developed.
		
		The strongly intensive variable $\Sigma(n_F,\ n_B)$ was calculated for different energies for two values of the width of the observation rapidity windows as a function of the distance between centers of these windows. It has been shown that $\Sigma(n_F,\ n_B)$ increases with both initial energy of pp collision and centrality.
		
		As is clear from formula (\ref{sigma rewritten}), in both cases this is caused by the growth of the proportion of string clusters with a larger number of merged strings. Recall that, as it was shown in \cite{Vechernin18,Andronov2019,Vechernin19,Belokurova Vechernin 2019}, the value of the variable $\Sigma(n_F,\ n_B)$ depends only on the properties of sources and the proportion in which they are formed in a collision.
		 
		Using the obtained MC simulation results the scaled variance $\omega_n$ and robust variance $R_n$ in pp collisions for different energies and for different width of the observation rapidity window were also calculated.


\begin{thebibliography}{99}
		\bibitem{col1}A.B.Kaidalov, Phys. Lett. B  \textbf{116:6}, 459-463 (1982)
		\bibitem{col2}A.B.Kaidalov, K.A.Ter-Martirosyan, Phys. Lett. B,  \textbf{117}, 247 (1982)
		\bibitem{col3}A.Capella, U.Sukhatme, Chung-I Tan, J.Tran Thanh Van, Phys. Lett. B \textbf{81:1}, 68-74 (1979).
		\bibitem{col4}A.Capella,U.Sukhatme, Chung-I Tan, J. Tran Thanh Van, Phys. Rep.  \textbf{236}, 225 (1994)
		\bibitem{int1}T.S. Biro, H.B. Nielsen, J. Knoll, Nucl. Phys. B \textbf{245}, 449 (1984)
		\bibitem{int2}A. Bialas, W. Czyz, Nucl. Phys. B \textbf{267}, 242 (1986)
		\bibitem{model1}M.A. Braun, C. Pajares, Phys. Lett. B \textbf{287}, 154 (1992)
		\bibitem{model2}M.A. Braun, C. Pajares, Nucl. Phys. B \textbf{390}, 542 (1993)
		\bibitem{uni distr 0}V.V. Vechernin and R.S. Kolevatov, Vestn. Peterb. Univ. Ser. 4: Fiz. Khim., No. 2, 12–23 (2004); arXiv:hep-
		ph/0304295v1
		\bibitem{uni distr 1}V.V. Vechernin and R.S. Kolevatov, Vestn. Peterb. Univ. Ser. 4: Fiz. Khim., No. 2, 11–27 (2004); arXiv:hep-
		ph/0305136v1
		\bibitem{uni distr 2}M.A. Braun, R.S. Kolevatov, C. Pajares, V.V. Vechernin, Eur. Phys. J. C \textbf{32:4}, 535-546 (2004); arXiv: hep-ph/0307056
		\bibitem{PoS2012}	Vechernin V.V., Lakomov I.A., The dependence of the number of pomerons on the impact parameter and the long-range rapidity correlations in pp collisions, PoS (Baldin ISHEPP XXI) 072.
		\bibitem{ALICE15} J.~Adam, D. Adamova, M.M. Aggarwal et al. [ALICE Collab.],\emph{JHEP} \textbf{05} (2015), 097, arXiv:1502.00230.
		\bibitem{NPA15} V.~Vechernin, \emph{Nucl. Phys. A} \textbf{939} (2015), 21-45, arXiv: 1210.7588.
		\bibitem{Pogosayn12rep}		M. Poghosyan. Inelastic and diffraction dissociation cross-sections in proton-proton collisions with ALICE, LHC Seminar, CERN, 09.10.2012, https://indico.cern.ch/event/211168/.
		\bibitem{Vechernin18} V.V. Vechernin, EPJ Web Conf. 191, 04011 (2018).
		\bibitem{Andronov2019}	E. Andronov, V. Vechernin, Eur. Phys. J. A 55 (2019) 14.
		\bibitem{Vechernin19} V. Vechernin, E. Andronov, Universe 5, 15 (2019).
		\bibitem{Gorenstein11} M.I. Gorenstein, M. Gazdzicki, Phys. Rev. C 84, 014904 (2011).
		\bibitem{AndronovTMPh15}. E.V. Andronov, Theor. Math. Phys. 185, 1383 (2015).
		\bibitem{TOTEM}		G. Antchev et al. (The TOTEM Collaboration), EPL, 96 (2011) 21002.
		\bibitem{CMS} V. Khachatryan et al. (CMS Collaboration) PRL 105, 022002 (2010).
		\bibitem{ALICE7n} K. Aamodt et al. (ALICE Collaboration),  Eur. Phys. J. C 68 (2010) 345.
		\bibitem{TOTEM13} G. Antchev et al. (The TOTEM Collaboration),  Eur. Phys. J. C 79 (2019) 103.
		\bibitem{Belokurova Vechernin 2019} S. N. Belokurova, V. V. Vechernin, Theoretical and Mathematical Physics, 200(2): 1094–1109 (2019).
	\end{thebibliography}
\end{document}